# Coronavirus Geographic Dissemination at Chicago and its Potential Proximity to Public Commuter Rail

Peter Fang

Perceptiongraph LLC, Glenview, IL

**Abstract**: The community spread of coronavirus at great Chicago area has severely threatened the residents' health, family and normal activities. CDC daily updates on infected cases on County level are not satisfying to address public's concern on virus spread. On March20th, NBC5 published case information of 435 coronavirus infections. The data is relative comprehensive and of high value for understanding on the virus spread patterns at Chicago. Data engineering, natural language processing and Google map technology are applied to organize the data and retrieve geographic information of the virus. The analysis shows community spread in Chicago areas has a potential proximity relation with public commuter rail. Residents nearby major public commuter rails need limit outdoor activities during the outbreak and even the post-peak time.

**Introduction**: The panic among Chicago residents have escalated rapidly as lacking basic knowledge of virus community spread regarding key questions on the geographic distribution, pollution concentration, self-protection, and forecast on contingency. Solutions should be able be found as substantial reliable data are available. CDC so far didn't disclose sufficient information to the public may have some concerns, especially data collection, analysis and modeling takes time. The pain point is that public needs to know more accurate information about the virus at Chicago so to manage the crisis and reduce the risk of personal health and the prohibition of their economic activities.

The needs on crisis manage lead to a comprehensive report of 435 virus report from the NBC 5 Chicago up to March 20th. The dataset doesn't contain patient personal information but provide important county, city, and travel history data. The data are of extremely high value as it provides sufficient information for scientists to model the virus dissemination patterns in Chicago areas because of the travel history can differentiate the patient whether being affected from travel with community infection. Secondly, the city data significantly narrow down the geographic positions, contrasting to current status that only county data are provided, for example a data has city information in the Cook County is more accurate than the County in terms of the spatial resolution.

**Methods and hypothesis**: To extract the virus dissemination patterns from the NBC 5 reports are challenging. First, 200 city data are missing and the City of Chicago is too big to decide a relatively accurate position of the virus. Statistics and natural language processing are used in this work to solve the problem. Statistics can determine the most likely city in a County where the virus is reported. If the city data for a given case is missing, we can impute the missing one with the most likely city name. Cases in the City of Chicago are hard to determine the locations. Fortunately, the travel history data from the NBC 5 have patients' working address or relevant addresses such as condominium or gym. It is safe to hypothesize a 60-70% probability that the patients are infected in areas close to their working addresses if not travel related or directly contact other known cases.  Therefore, the addresses information in the history data are extracted and used in this work.

In addition to the missing city data, accurate differentiation between travel related and community infection is also hard to solve because that the history data is a log file recorded by different agents using non-uniform formats, documentation methods, and writing. The problem is solved using regex programming and a data ETL (Extract-transform-load) pipeline.

The most challenge is how to deliver a reasonable understanding on the virus dissemination pattern without misinformation and misleading conclusions.

**Tools and Technology**:  Python programming, including pandas and regex, are used to clean the unstructured data. Natural language processing package Spacy is used to extract address or position information from the travel history data. Python package Gmaps is used to display data. Google map API and its function of transit route view are used to for data visualization.

**Discussion**:

The travel related or communities infected are separated by data engineering pipeline.  Below I show samples of travel related cases and communities infected cases in different tables.

| | data_annouced | COUNTY | CITY | histroy_result | histroy |
|---|---|---|---|---|---|
| **128** | 2/29/2020 | Cook | Unknown | travel | spouse of woman who also tested positive. likely exposed from travel to another state that had community transmission. released from northwest community hospital to home isolation. |
| **130** | 3/5/2020 | Cook | Chicago | travel | acquired virus on the grand princess cruise ship (where several other passengers also tested positive); employed as a special education aide at vaughn occupational high school in chicago |
| **135** | 3/9/2020 | Cook | Chicago | travel | california resident who traveled to illinois |
| **136** | 3/9/2020 | Cook | Chicago | travel | returned earlier in march from an egyptian cruise which had been linked to other cases |
| **139** | 3/15/2020 | Sangamon | Springfield | travel | florida resident who'd traveled to springfield. she was the first confirmed coronavirus case in the county. she died on march 19, 2020. |
| **142** | 1/24/2020 | Cook | Chicago | travel | returned from wuhan, china in mid-january; discharged from st. alexius medical center in hoffman estates in early february; finished recovery at home |
| **143** | 1/30/2020 | Cook | Chicago | travel | spouse of woman, who had also tested positive, who had returned from wuhan, china in mid-january; discharged from st. alexius medical center in hoffman estates in early february; completed recovery at home. first recorded human-to-human transmission in the u.s. |
| **144** | 3/5/2020 | Cook | Unknown | travel | vanderbilt university student from the chicago area, who flew into o'hare in early march, after traveling to italy |
| **147** | 3/15/2020 | Champaign | Unknown | travel | was in contact with someone who had traveled to italy. as of 3/15: at home in isolation and recovering well. no association with u of i |
| **149** | 3/19/2020 | Jackson | Unknown | travel | possibly exposed to the virus during travel to another state. as of 3/19: at home and insolation. first reported case in jackson county |

Table I: Samples of travel related cases.  The 'History' column shows the original data describing patients travel history.  'History_result' is a classification on the contents of the 'History' column whether the cases are travel related.

| | data_annouced | COUNTY | CITY | histroy_result | histroy |
|---|---|---|---|---|---|
| 145 | 3/10/2020 | McHenry | Unknown | community | now released from isolation. no known history of travel to an affected area; no connection to a known case of covid-19. |
| 146 | 3/12/2020 | McHenry | Unknown | community | now released from isolation. |
| 148 | 3/18/2020 | Cook | Oak Park | community | first diagnosed case in oak park. as of 3/18: at home, in isolation |
| 150 | 3/19/2020 | LaSalle | Unknown | community | no history of travel or contact with an existing case. as of 3/19: recovering at home in isolation. first confirmed case in lasalle county |
| 151 | 3/18/2020 | Peoria | Unknown | community | as of 3/17: at home and in isolation |
| 152 | 3/18/2020 | Peoria | Unknown | community | as of 3/17: at home and in isolation |
| 153 | 3/18/2020 | Will | Joliet | community | student at joliet junior college. as of 3/18, was released from a hospital in cook county |
| 154 | 3/19/2020 | St Joseph | Unknown | community | had exposure history to a known covid-19. as of 3/19: observing self-isolation as instructed |
| 155 | 3/18/2020 | St. Joseph | Unknown | community | as of 3/18: has not required hospitalization; observing self-isolation |
| 156 | 3/15/2020 | Clinton | Unknown | community | non-resident staying in clinton county; no known international travel. as of 3/18: self-quarantined; seems to be doing well |

Table II: Samples of communities infected cases. 'History' column is the original data describing patients travel history. 'History_result' column briefs the 'History' column as commute spread. If not explicitly noting a travel related, cases are considered as community spread. It is pointing out here that cases in the original data such as "no details reported by public health officials or in news stories" are not useful in our analysis and were dropped out. Around 160 data are used in the current analysis. 160 data is a smaller sample set but of high quality that sufficiently provides valuable information on certain extent.

After separating travel-related cases with community infected cases, the two groups of data are visualized on maps using Google map API. Only Chicago areas are focused as most cases are concentrated in Chicago and its suburbs. The solid green circle in Graph I represents travel related cases or patients who have close contacts with travel cases, including people from China, Italy, Cruise, California or other states in US. The red solid circles are the community infected cases. Because of the size of each red circle, many cases overlap on each other if their locations are close, for example, Willowbrook nursing center has 46 cases up to March 20$^{th}$. From the visualization, many cases happened in Downtown Chicago, which is not very surprising as flux of people in and out downtown daily. An interesting feature in Graph I is that community spread cases (red points) seems loosely connect with the green circles where most of them are at the north-west side Chicago such as Arlington Height. The loose connection between two group data doesn't support community dissemination comes from north-west side Chicago. One striking feature of the graph is that the red points stretch from downtown to northbound Chicago within a confined stripe-like pattern, indicating virus spread taking advantage of a special media. A most straightforward bet is the public transit. To valid the hypothesis, four public transit lines are visualized on the same graph. MD-N line and UP-N line almost superimpose on the confined stripe-like pattern of the virus dissemination. The coincidence suggests a potential proximity relation between virus spread in Chicago with the public commuter rails. If the proximity true, it can explain the outbreak of the Willowbrook nursing center where is close to BNSF line. Some stuffs, visitors or residents may commute to downtown Chicago via the line recently.

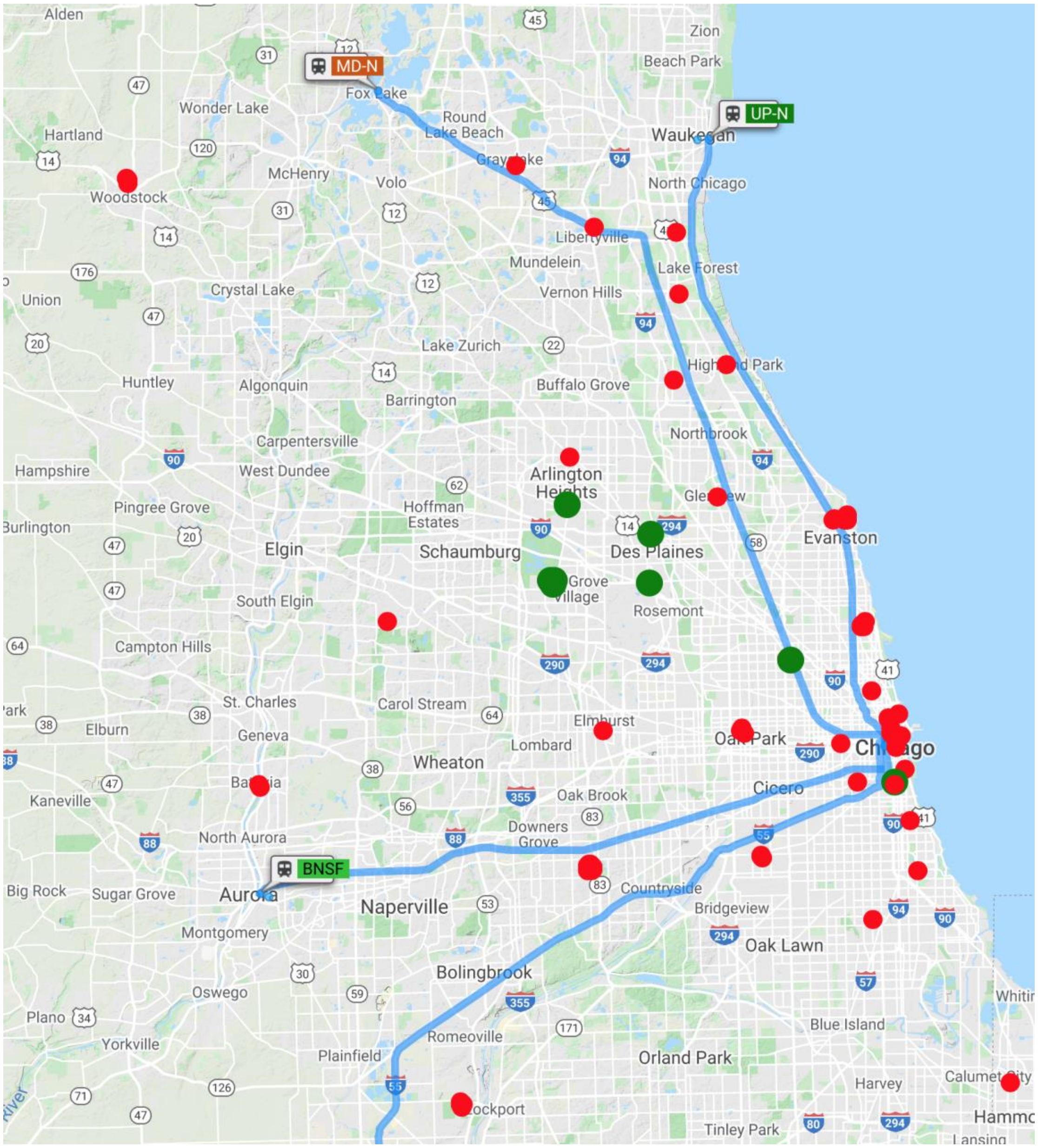

Graph I. Communities spread of coronavirus in Chicago may have a proximity effect with the public commuter rails. Solid red points are reported cases due to community spread. The green points represent cases introduced from abroad or other states in US. The blue lines are public commuter rails of Chicago.

**Conclusion**:

The analysis shows community spread of coronavirus in Chicago areas has a potential proximity relation with public commuter rails. Residents nearby major public commuter rail need limit outdoor activities during the outbreak and even the post-peak time.

The proximity concept introduced in this paper cannot be generalized to other major metro areas where virus spread patterns could be significantly different. The concept may not valid anymore if more migrated cases hit Chicago in a short time.